\documentclass[aps,pra, reprint]{revtex4-1}

\usepackage{amsmath,amssymb}
\usepackage[english]{babel}
\usepackage{graphicx}
\usepackage{dcolumn}
\usepackage[squaren, cdot]{SIunits}
\usepackage{hyperref}
\usepackage{color}
\usepackage{braket}
\renewcommand{\vec}{\mathbf}
\newcommand{\eq}[1]{Eq.~\eqref{#1}}
\newcommand{\fig}[1]{Fig.~\ref{#1}}
\newcommand{\tab}[1]{table~\ref{#1}}
\newcommand{\aopph}[1]{{\hat{a}_{#1}^{\phantom \dag}}}
\newcommand{\aop}[1]{{\hat{a}_{#1}}}
\newcommand{\aopd}[1]{{\hat{a}_{#1}^\dag}}
\newcommand{\bop}[1]{{\hat{b}_{#1}}}
\newcommand{\bopph}[1]{{\hat{b}^{\phantom \dag}}_{#1}}
\newcommand{\bopd}[1]{{\hat{b}^\dag}_{#1}}
\newcommand{\ald}[2]{\hat{\alpha}^{\dagger #1}_{#2}}
\newcommand{\al}[2]{\hat{\alpha}^{#1}_{#2}}
\newcommand{\betd}[2]{\hat{\beta}^{\dagger #1}_{#2}}
\newcommand{\bet}[2]{\hat{\beta}^{#1}_{#2}}

\begin{document}
% title of paper
\title{Quantum simulators by design - many-body physics in reconfigurable 
arrays of tunnel-coupled traps}

% authors and affiliations
\author{M. R. Sturm}
\email{martin.sturm@physik.tu-darmstadt.de}
\author{M. Schlosser}
\author{R. Walser}
\author{G. Birkl}
\affiliation{Institut f\"ur Angewandte Physik, Technische Universit\"at Darmstadt, 64289 Darmstadt, Germany}

% the date
\date{\today}

\begin{abstract}
We present a novel platform for the bottom-up construction of
% tunneling based 
itinerant many-body systems: ultracold atoms transferred from a  Bose-Einstein 
condensate into freely configurable arrays of micro-lens generated 
focused-beam dipole traps.
This complements traditional optical lattices
and gives a new quality to the field of two-dimensional 
quantum simulators. 
The ultimate control of topology, well depth, atom number, and interaction strength is matched by sufficient tunneling.
We characterize the required light fields, derive the Bose-Hubbard 
parameters for several alkali species, investigate the  loading 
procedures and heating mechanisms.
To demonstrate the potential of this approach, we analyze coupled annular Josephson contacts exhibiting many-body resonances.
\end{abstract}

\maketitle 

\section{Introduction}

With the realization of the superfluid/Mott-insulator transition in optical 
lattices \cite{Greiner2001}, ultracold atoms in periodic optical potentials 
have become a versatile toolbox for the study of quantum many-body physics 
\cite{Jaksch1998,Lewenstein2012}. The recent invention of quantum gas 
microscopes \cite{Bakr2009,Sherson2010} extended this work to the 
observation of local properties, such as spreading of correlations 
\cite{Cheneau2012}, dynamics of spin impurities \cite{Fukuhara2013}, quantum 
walks 
\cite{Preiss2015a}, and entanglement entropy \cite{Islam2015}. Although 
these setups allow for local modifications of the potential surface and the 
atom properties using holographic masks \cite{Bakr2009}, spatial light 
modulators \cite{Fukuhara2013,Preiss2015a} or tightly focused laser beams 
\cite{Weitenberg2011}, the underlying lattice structure remains periodic.

Complementing this approach, recent experiments with tightly focused optical 
tweezers demonstrated efficient trapping and cooling of single atoms to 
their vibrational ground state \cite{Kaufman2012,Thompson2013}. 
In combination with acousto-optic deflectors and 
spatial light modulators (SLM) this approach has been extended to few-well configurations showing tunnel-coupling \cite{Kaufman2014,Murmann2015} on one side and the deterministic preparation of larger defect-free 1D and 2D arrays with spacing too large for tunneling on the other side \cite{Barredo2016, Endres2016, Kim2016}.

\begin{figure}
	\includegraphics[scale=1]{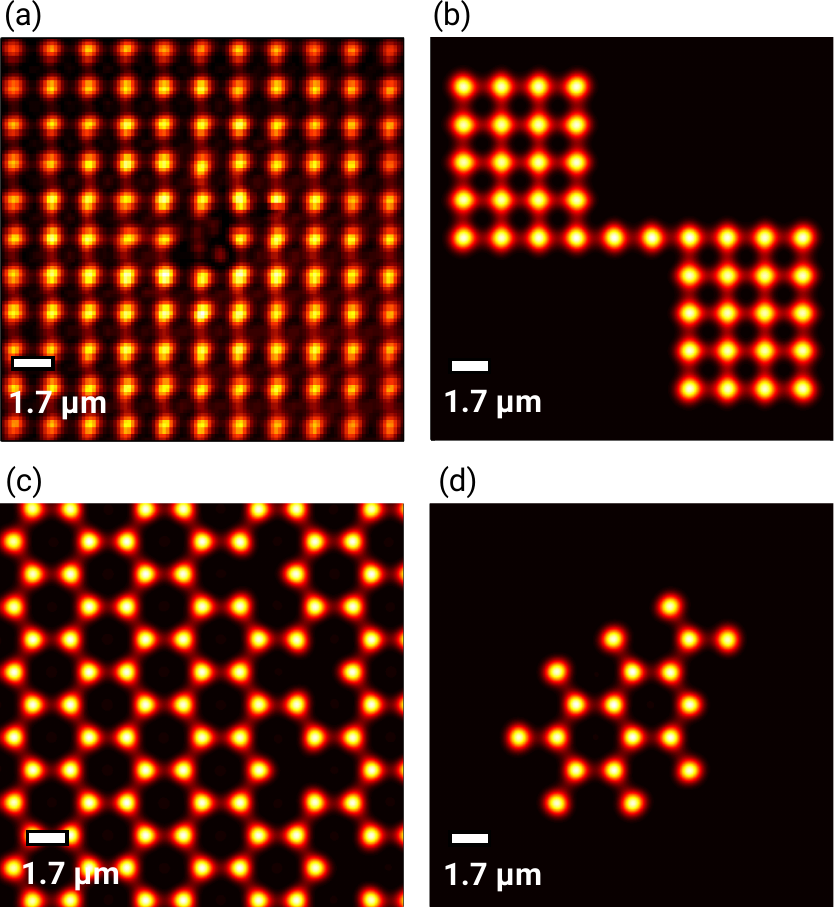}
	\caption{(color online) Optical intensities for a set of trap configurations for a novel type quantum simulator. Optical intensity $\mathcal{I}(\vec{x})$ of planar lattice-type geometries of focused-beam dipole traps separated by $\unit{1.7}{\micro m}$: (a) measured intensity in a square lattice with designed lattice defect, (b) simulated intensity for an atomtronic diode, (c)~simulated graphene-like lattice with several point defects on the right,  (d)~simulated molecular structure of quinolones, a family of antibiotics.}
	\label{fig:opticalPotential}
\end{figure}

In this article, we introduce a novel experimental platform for quantum many-body physics by combining the advantages of the above approaches: We create versatile  patterns of optical microtraps (see Fig.~\ref{fig:opticalPotential}) with comprehensive single-site control using microlens arrays (MLA) \cite{Birkl2001,Kruse2010,Schlosser2011,Schlosser2012} in combination with spatial light modulators. Achievable trap parameters ensure to enter the tunneling based many-body regime for large-scale systems with full single-site control. Each site corresponds to an individual cross-talk-free diffraction-limited laser spot. Arbitrary periodic and non-periodic 2D potential surfaces on a micrometer scale with dynamic control of the trap parameters can be implemented. For atoms transferred from a Bose-Einstein condensate (BEC) into this optical potential defect free occupation of each site is automatic. In addition, we can dial-up tunneling rates, on-site interactions, and trap frequencies for each site individually or in parallel in order to cross from the superfluid to the strongly interacting many-body regime. This gives a new quality to the experimental study of, e.g., transport phenomena, finite-size effects, crystal defects, quasi-periodic structures, disorder, frustration, 2D magnetism, and their dynamic control. Consequently, this bottom-up approach significantly extends the successful top-down approach to quantum simulation using 'traditional' optical lattices.

In contrast to interfering laser waves, our architecture provides direct single-site control while beeing phase insensitive and structurally robust. Compared to state of the art holographic trap arrays generated by phase modulating spatial light modulators \cite{Barredo2016, Kim2016}, it accesses the tunneling regime and omits pixelation constraints imposed to trap spacing, homogeneity and system size. With regard to potentials generated by acousto-optics through multi-tone synthesis \cite{Endres2016} or time-averaging \cite{Zimmermann2011}, it is scalable and avoids additional heating.

The paper is organized as follows: In section \ref{sec:setup} we give a description of our setup and show measurements and simulations of the light field for several geometries. In section \ref{sec:rings} we analyze a prototypical application consisting of two weakly coupled ring lattices. In section \ref{sec:feasibility} we study the experimental feasibility of our approach by characterizing the light fields using measurements and simulations, computing the Bose-Hubbard parameters, and analyzing the impact of light scattering. Finally, in section~\ref{sec:outlook} we conclude and provide an outlook to additional applications.

\begin{figure}
	\includegraphics[width=\linewidth]{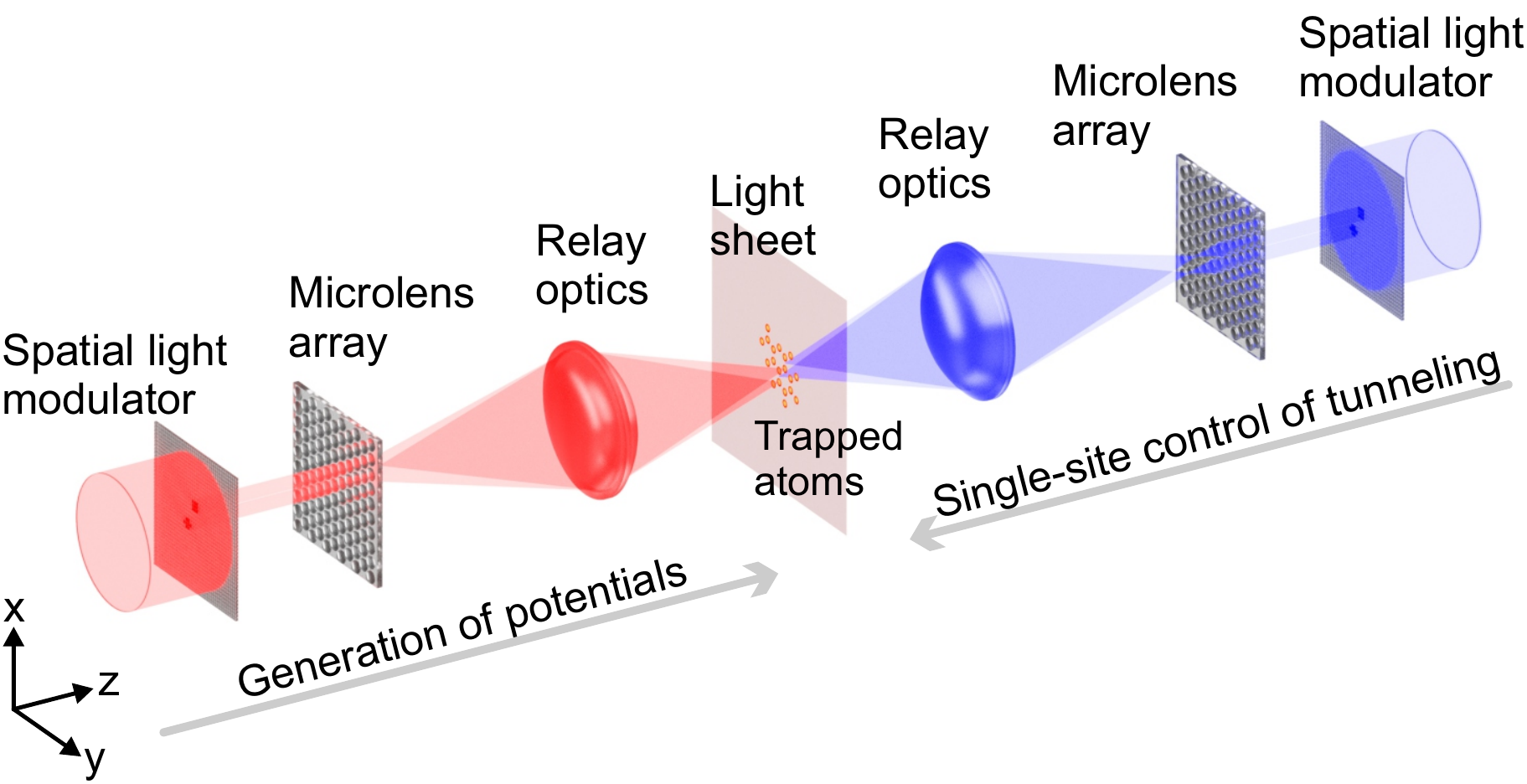}
	\caption{(color online) Experimental setup for flexible creation of 2D potential geometries (red from left) and site-selective control of the tunneling rates (blue from right). Both light fields are generated by a combination of  microlens arrays and a spatial light modulators controlling the illumination of each lenslet. The resulting focal planes are reimaged into a vacuum chamber by two demagnifying relay lens systems. For out-of-plane confinement of the atoms via $V_{\parallel}$ an additional light sheet is applied in the $x$-$y$ plane.}
	\label{fig:setup}
\end{figure}

\section{Optical potential and experimental setup}
\label{sec:setup}

\begin{figure}
	\includegraphics[scale=1]{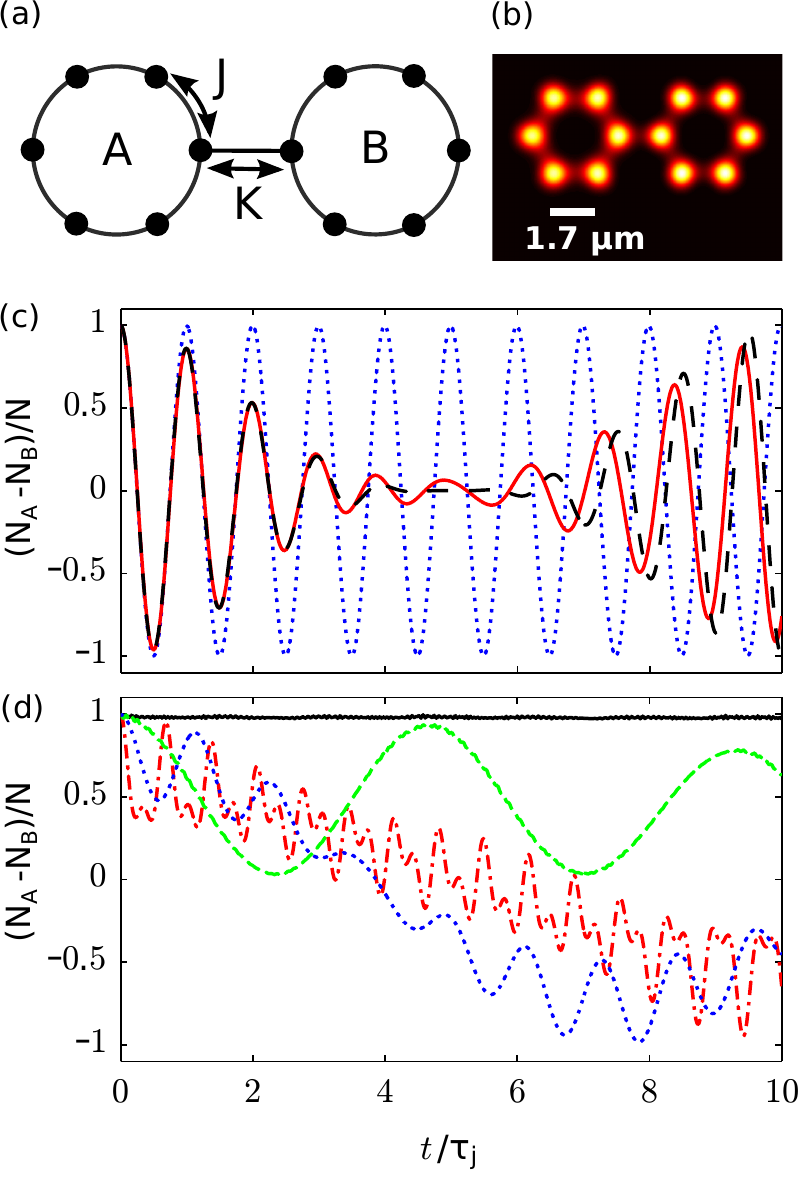}
	\caption{(color online) Coupled Josephson rings. (a) Josephson point contact of two ring lattices with $M=6$ sites per ring, (b) simulated optical intensity $\mathcal{I}(\bf{x})$ using a hexagonal MLA, (c) and (d) simulation of the population inversion subject to the Bose-Hubbard Hamilton operator with $N=4$ particles for $K=0.1 J$: $U=0$ ((c), dotted blue), $U=0.02 J$ ((c), solid red) compared to the perturbation theory result of \eq{eqn:imbalance} ((c), dashed black), $U=J$ ((d), solid black), $U=3 J$ ((d), dash-dotted red), $U=5.07 J$ ((d), dashed green), and $U=11.7 J$ ((d), dotted blue).}
	\label{fig:CoupledRings}
\end{figure}

The intensity distributions of selected configurations is depicted in figure~\ref{fig:opticalPotential}: (a) square lattice with a defect implemented by site-selective control with the SLM, (b) atomtronic diode \cite{Pepino2009} as a pinboard for various atomtronics devices \cite{atomtronics}, (c) hexagonal lattice as in graphene, including defects, (d) or even complex organic molecules \cite{Luehmann2015} like quinolones, a family of antibiotics. A schematic representation of the experimental setup is shown in figure~\ref{fig:setup}. Two subsystems consisting of a microlens array and a spatial light modulator each are combined to create flexible 2D potential geometries (red from left in \fig{fig:setup}) and site-selective control of the tunneling rates (blue from right in \fig{fig:setup}). The intensity distribution $\mathcal{I}(\mathbf{x})$ is proportional to the optical dipole potential experienced by the atoms:
\begin{math}
V(\mathbf{x})=\mathcal{D}^2\mathcal{I}(\mathbf{x})/\hbar \delta,
\end{math}
with $\mathcal{D}$ being the effective atomic dipole and $\delta$ the detuning \cite{cohentannoudji92,Steck2007}. We compose the total potential 
\begin{math}
V(\mathbf{x}) = V_\bot (\mathbf{x}) + V_\parallel (\mathbf{x}), 
\end{math}
which confines atoms in a planar lattice-like geometry. Here, $V_\bot$ is a microlens-generated 2D optical dipole trap array and the light sheet potential $V_\parallel$ provides out-of-plane confinement. In this novel platform, we can implement many-body states with defined particle number in freely configurable geometries. The intensity distribution shown in Fig. \ref{fig:opticalPotential} (a) is obtained by a $65$-fold demagnification of the focal plane of a fused-silica MLA with $\unit{110}{\micro\meter}$-period using relay optics with $\text{NA}=0.68$ giving $d=\unit{1.7}{\micro m}$ and $w_0=\unit{0.71}{\micro m}$. Linearly polarized light with a wavelength of $\lambda=\unit{780}{nm}$ and a liquid-crystal-based SLM for single-site control are used. Figure 1 (b) to (d) and Fig. \ref{fig:CoupledRings} (b) depict simulated intensity distributions for an equivalent diffraction-limited optical system and $\lambda=\unit{1064}{nm}$ using optical design software. This results in $d=\unit{1.7}{\micro m}$ and $w_0=\unit{0.74}{\micro m}$. The value of the trap spacing is the result of an optimization of the trade-off between maximum tunneling strength and cross-talk-free single-site control.

%\textcolor{blue}{
%Compared to other experimental techniques our approach has the following advantages and disadvantages. The advantages compared to optical lattices have been discussed in the introduction. On the downside, our approach results in larger trap spacings and therefore slower dynamics (cf. section \ref{sec:BHP}). Compared to holographic trap arrays generated by SLMs \cite{Nogrette2014, Zupancic2016}, it is less flexible in generating arbitrary potential landscapes. On the upside, the light field is structurally more robust, since its structure is determined by the MLA. For the same reason it should also be easier to generate large homogeneous trap arrays. In contrast to arrays of optical tweezers formed by acousto-optic modulators \cite{Zimmermann2011}, our approach allows for a wider range of geometries and can be scaled to larger trap numbers. In addition it avoids heating arising from the use of time-averaged potentials \cite{Zimmermann2011}.
%}

\section{Coupled Josephson rings}
\label{sec:rings}
An example for the potential of this approach is a Josephson point contact between two one-dimensional ring lattices (Fig. \ref{fig:CoupledRings}). This paves the road from the investigation of double-well dynamics \cite{Albiez2005, Levy2007, Kaufman2014, Murmann2015} to the study of many-body physics of more complex systems, i.\thinspace{}e., quantum dots or superfluid nuclei \cite{Kleber1971}, allowing for the investigation of multi-particle resonances in a fully controlled fashion. In the configuration of Fig. \ref{fig:CoupledRings} (a), ring $A$ constitutes a 1D periodic, $M$-site lattice with a Bose-Hubbard Hamilton operator
\begin{equation}
\hat{H}_A = -J \sum_{\langle i j \rangle} (\aopd{i} \aopph{j} 
+ \aopd{j} \aopph{i})+ 
\frac{U}{2} \sum\limits_{i=0}^{M-1} \aopd{i}\aopd{i}\aopph{i}\aopph{i},
\label{eqn:BHH}
\end{equation}
where $J$ is the tunneling energy, $U$ is the on-site interaction strength, $\aop{i}$ is the bosonic particle annihilation operator for site $i$, and $\langle i j \rangle$ denotes the summation over nearest-neighbor pairs. For ring $B$, $\hat{H}_B$ is obtained analogously by substitution $\aop{i} \rightarrow \bop{i}$. At the site $i=0$, we introduce a weak link of strength $K < J/M$ between the rings. Thus, the total energy is $\hat{H}=\hat{H}_A+\hat{H}_B-K (\aopd{0}\bopph{0} + \bopd{0} \aopph{0})$. Experimentally, the weak link can be tuned by increasing the spacing between both rings or by applying a tightly focused blue-detuned beam increasing the potential barrier between the rings (cf. 'blue side' of \fig{fig:setup}). After loading $N$ atoms into the interacting ground state of the isolated ring $A$, we connect the junction and observe oscillations in the population inversion $\zeta=(N_A-N_B)/N$ between both rings, with $N_A = \sum_i \langle \aopd{i}\aopph{i} \rangle $ and $N_B=\sum_i \langle \bopd{i}\bopph{i} \rangle$.

We compute the dynamics for arbitrary interaction strength by exact numerical diagonalization  \cite{Sidje1998}. In Fig. \ref{fig:CoupledRings} (c), (d), we show the population oscillations for $M=6$ sites and $N=4$ particles for different on-site interaction strengths. For $NU\ll J$, the non-interacting ground-states of the isolated rings mostly define the dynamics as shown in Fig. \ref{fig:CoupledRings} (c) for $U=0$ (c, blue) and $U=0.02J$ (c, red), respectively. A two-site Bose-Hubbard model with rescaled interaction energy $u=U/M$ and tunneling strength $k=K/M$ effectively captures the physics. Within first order perturbation theory, the population inversion reads (cf. appendix \ref{sec:appendix})
\begin{equation}
\zeta(t)= \cos{(\omega_{j} t)} \cos^{N-1} (\pi t/\tau_r) 
\stackrel{t<\tau_c}{\longrightarrow}
\cos{(\omega_j t)} e^{-\frac{t^2}{\tau_c^2}} .
\label{eqn:imbalance}
\end{equation}
For weak on-site interaction ${u\ll k}$, the Josephson oscillations with period $\tau_j=2\pi/\omega_j=h /2 k$ collapse on the time scale $\tau_c=\tau_r\sqrt{2/(\pi^2 (N-1))}$ and revive subsequently at $\tau_r= h /u$ (dashed black line in Fig. \ref{fig:CoupledRings} (c)). In the limit $u=0$, these Josephson oscillations persist indefinitely. Both results agree with exact diagonalization, as shown in Fig. \ref{fig:CoupledRings} (c).

\begin{figure}
	\includegraphics[scale=1]{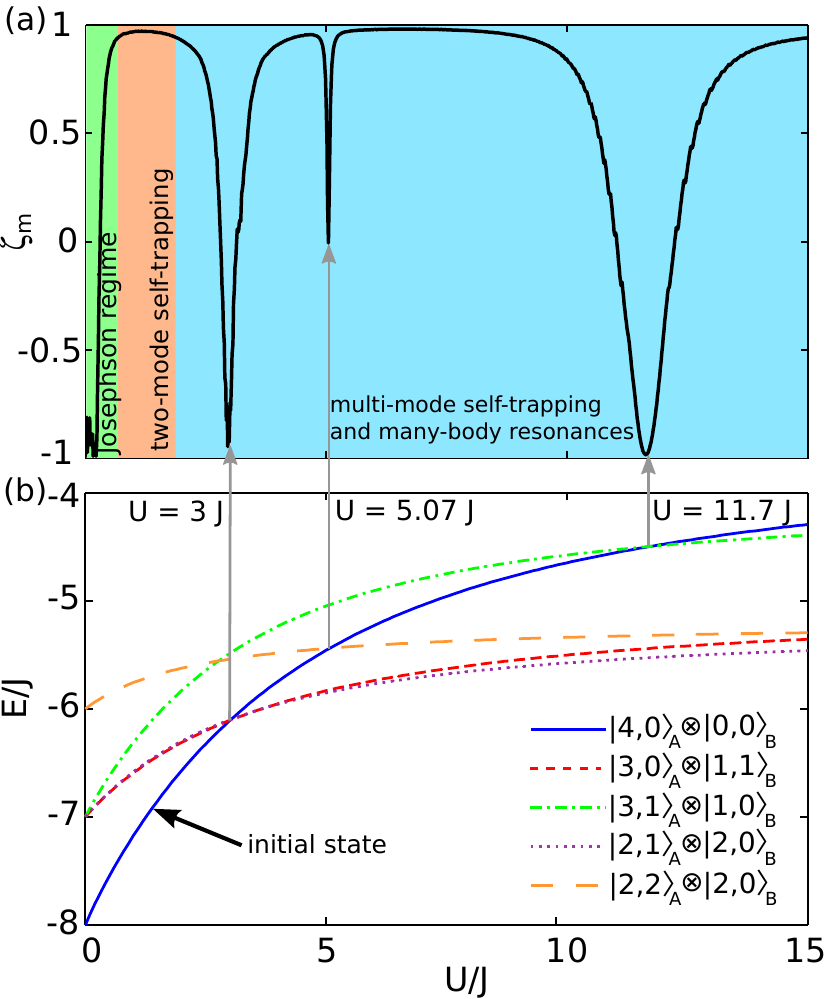}
	\caption{(color online) Self-trapping and many-body resonances. (a) Minimal value $\zeta_m(U)$ of the population inversion between two coupled ring lattices for $M=6$ sites and $N=4$ particles  versus the interaction strength $U$ exhibiting many-body resonances. They correspond to the population oscillations shown in Fig. \ref{fig:CoupledRings} (c,d). (b) Eigenenergies of  this system for different distributions of the particles between the two rings. These energies correspond to states of the form $|N_A,j\rangle_A \otimes |N-N_A,j'\rangle_B$.}
	\label{fig:CoupledRingsResonances}
\end{figure}

For larger interaction strengths (Fig. \ref{fig:CoupledRings} (d)), the rich many-body dynamics of the coupled rings moves into focus: For $U\approx J$, the system enters the self-trapping regime and the population remains trapped in ring $A$ (d, black) as in a structureless double well potential \cite{Albiez2005}. 

However, the salient features are pronounced many-body resonances with significant population transfer for particular interaction strengths deep in the self-trapping regime (d, red, green, blue). To quantify the magnitude of the population transfer, we define $\zeta_m(U)$ as the minimal value of the population inversion in the interval $0<t<10 \, \tau_j$. Figure \ref{fig:CoupledRingsResonances} (a) shows $\zeta_m(U)$ as a function of $U$, revealing the Josephson and the self-trapping regimes, as well as the three pronounced many-body resonances of Fig. \ref{fig:CoupledRings} (d). These can be explained from the energy spectrum $\mathcal{E}_{j}(N,U)$ and the eigenstates $|N,j \rangle$ of isolated rings as a function of particle number $N$ and the interaction strength $U$ (cf. Fig. \ref{fig:CoupledRingsResonances} (b)). Here, $j$ is an integer with $-M/2 < j \le M/2$ corresponding to the quasi momentum $q_j=2 \pi j/(M d)$ enumerating the degenerate eigenvalues ($\mathcal{E}_{j}=\mathcal{E}_{-j}$) in ascending order with $|j|$. If for a given interaction strength the energy of the initial state $|N,0\rangle_A\otimes |0,0\rangle_B$ coincides with the energy $E$ of another state $|N_A,j\rangle_A\otimes |N_B=N-N_A,j'\rangle_B$, i.\thinspace{}e.
\begin{equation}
E\equiv \mathcal{E}_{j}(N_A,U)+\mathcal{E}_{j^\prime}(N_B,U)
=\mathcal{E}_{0}(N_A=N,U),
\end{equation}
we observe a resonance in the population transfer. In our system this will be accessible to direct experimental observation. A comparable effect has been predicted in the mean-field limit of Josephson junctions where resonant coupling to higher modes facilitates population transfer at large interaction strengths \cite{Julia-Diaz2010}. Further, interaction-induced tunneling resonances in tilted optical lattices have been observed \cite{Meinert2016}.

\section{Experimental feasibility}
\label{sec:feasibility}
Can this novel platform for many-body physics work? To give evidence for this, we analyze the reciprocal relations between large, optically accessible trap separations $d$ allowing for single-site addressability on one hand, and sufficiently high tunneling rates on the other hand. These apparently conflicting conditions depend themselves on the  in-plane  and out-of-plane optical potentials, $V_{\perp}$ and $V_{\parallel}$ respectively, which in addition affect the on-site interaction energy $U$. 

\subsection{Light field}

\begin{figure}
	\includegraphics[scale=1]{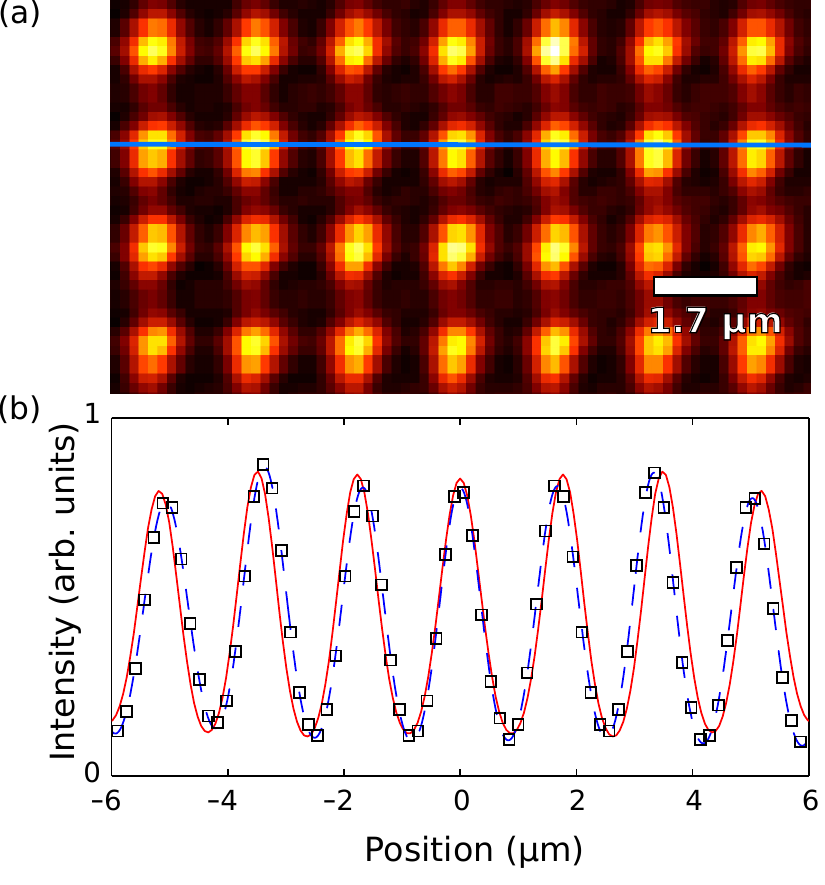}
\caption{(color online) Measured light field of a square lattice of optical microtraps: (a) central part of the intensity distribution; (b) section of the intensity distribution (black squares) along the horizontal line (a). The dashed blue line is a Gaussian fit to the experimental intensity distribution and the red line is the result of a simulation of the optical system.}
	\label{fig:comparison}
\end{figure}

In order to characterize the optical potential we analyze measurements and simulations of the light field. Figure \ref{fig:comparison} (a) shows the central part of measured intensity distribution using the setup described in section \ref{sec:setup}. In \fig{fig:comparison} (b) a cut of this intensity distribution (black squares) is compared to the result of a simulation of the full optical setup using commercial optical design software (red line) and a fit using the sum of seven Gaussian functions (dashed blue line). This analysis reveals that the in-plane potential is well approximated by a sum of Gaussian wavelets representing the trapping sites $\mathbf{R}_i=(X_i,Y_i)$
\begin{equation}
V_\bot (x,y) = - \sum_{\mathbf{R}_i} V_{0\perp}^{(i)} \,
e^{-2 \frac{(x-X_i)^2+(y-Y_i)^2}{w_{0\bot}^2}},
\label{eq:transversePotential}
\end{equation}
with local potential depths $V_{0\bot}^{(i)}$ controlled by the SLM. For a wavelength of $\lambda = \unit{1064}{nm}$ and a diffraction limited objective our simulations yield $d=\unit{1.7}{\micro m}$ and $w_{0\bot}$~=~\unit{0.74}{\micro m}. Further, the light sheet potential in the relevant region is given by $V_\parallel (z) = -V_{0\parallel} \exp (-2 z^2/w_{0\parallel}^2 )$, assuming an elliptical Gaussian beam with an out-of-plane width set to $w_{0\parallel}=$~\unit{2.5}{\micro m}. 

\subsection{Bose-Hubbard parameters}
\label{sec:BHP}
\begin{figure}
	\includegraphics[scale=1]{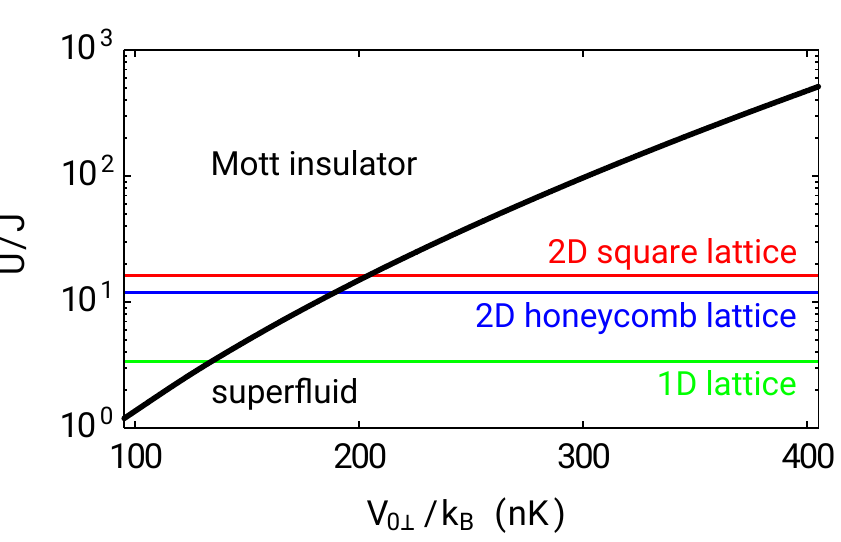}
	\caption{(color online) Bose-Hubbard parameters. Ratio of the Bose-Hubbard interaction energy $U$ and tunneling energy $J$ versus the optical potential depth $V_{0\bot}$ (thick black line) for $^{87}$Rb in a lattice with $d= \unit{1.7}{\micro m}$ and further  parameters given in table \ref{table}. The horizontal lines mark the predicted superfluid/Mott insulator phase transition for 1D and 2D geometries at unit filling \cite{Lewenstein2012,Luehmann2014}.}
	\label{fig:BHP}
\end{figure}

Using the parameters discussed in the preceding paragraph, we compute the interaction- and tunneling strengths from the eigenfunctions of the single particle Hamiltonian operator $\hat{H}_\text{sp} = \hat{\vec{p}}^2/(2m) + V(\hat{\vec{x}})$. Since the potential 
\begin{math}
%\label{eqv}
V(\mathbf{x}) \approx V_\bot (x,y) + V_\parallel (z), 
\end{math}
is approximately separable, the problem factorizes into an in-plane and an 
out-of-plane part. For the out-of-plane part $\phi(z)$, we choose the ground-
state of the corresponding 1D Schr\"odinger equation. For the in-plane part, 
we assume periodic boundary conditions and perform a band structure 
calculation \cite{Walters2013}, to obtain the Wannier functions $\varphi_i(
x,y)$ at lattice site $i$. We proceed by calculating the Bose-Hubbard 
parameters
\begin{align}
U &= \frac{4 \pi a_s \hbar^2}{m}\int  \varphi_i^4 (x,y) \phi^4(z) \; \text{d}
^3\vec{r}, \\
J &= \langle\varphi_i \phi | \hat{H}_\text{sp} |\varphi_j \phi\rangle, \\
\epsilon &=  \langle\varphi_i \phi | \hat{H}_\text{sp} |\varphi_i \phi\rangle.
\end{align}
Here, $U$ is the on-site interaction strength, $J$ is the tunneling parameter between adjacent sites $i$ and $j$, and $\epsilon$ is the local single-particle energy. Fig.~\ref{fig:BHP} depicts the results for $^{87}$Rb atoms in the state $|5^2 S_{1/2},F=1,m_F=-1 \rangle$ with scattering length $a_s=$~100.4~$a_0$ in 1D, 2D square, and 2D honeycomb lattices. In table~\ref{table} the relevant parameters for various bosonic alkali species are given at $U/J=10$. With reported life times in BECs in optical potentials well above \unit{10}{\second}, these results confirm that the superfluid/Mott-transition can indeed be reached for various lattice geometries at realistic experimental parameters. The accessible range of $U$ and $J$ is determined by limitations on the potential depth. We expect the lower limit at $V_{0\bot}=h^2/(2m d^2)$ as it was shown that the single-band Bose-Hubbard model ceases to be valid for traps shallower than that \cite{Pilati2012, Astrakharchik2016}. At this point $J$ is about $20$ times larger and $U$ about $3$ times smaller compared to the values given in table \ref{table}. The opposite limit of deep traps allows to effectively implement $U/J \to \infty$ due to the exponential suppression of $J$.

\subsection{Fluctuations of potential depth}
Experimental parameters, like the optical potential depth, are afflicted by 
spatio-temporal fluctuations. Thus, the parameters $U$, $J$, and $\epsilon$ 
acquire the uncertainties $\Delta U$, $\Delta J$, and $\Delta \epsilon$, 
respectively.  Fortunately, our MLA is inherently robust against temporal 
changes in spacing or shape of the traps. Therefore, we expect the temporal 
fluctuations of these quantities to be negligible. However, the depth of 
each trap is determined by the laser power that illuminates the respective 
microlens giving rise to temporal fluctuations of this quantity. Further, 
production tolerances of the MLA and aberrations of the demagnification 
optics may result in spatial fluctuation of the optical potential 
parameters. In the following, we will quantify the variations of the Bose-
Hubbard parameters due to fluctuations of the trap depths.

In order to compute the influence of the aforementioned fluctuations, we solve the 2D Schr\"odinger equation for a stochastic potential $V_\bot(x,y)$ of equation~(4). We draw the local trap depths as identically distributed normal variables $V_{0\bot}^{(i)}$ with mean $V_{0\bot}$ and standard deviation $\Delta V_{0\bot}$. The resulting potential is no longer periodic so instead of using band structure calculation we apply the Lanczos method to compute the lowest energetic eigenstates of the single-particle Hamilton operator $\hat{H}_\text{sp}$. We calculate maximally localized superpositions of the lowest energy eigenstates that correspond to the Wannier functions in the periodic case. In the absence of fluctuations, i.\thinspace{}e., $\Delta V_{0\bot}=0$, the systematic relative error of both methods is on the $10^{-3}$ level for all Bose-Hubbard parameters. For $\Delta V_{0\bot}>0$, we used $100$ samples to estimate the variance of the Bose-Hubbard parameters. For small fluctuations of the trap depth $\Delta V_{0\bot}/V_{0\bot} \ll 1$,  we find linear relations 
\begin{equation} 
\frac{\Delta J}{J} 
\approx C_{J} \frac{\Delta V_{0\bot}}{V_{0\bot}},\; \frac{\Delta U}{U} 
\approx C_{U} \frac{\Delta V_{0\bot}}{V_{0\bot}}, \; \frac{\Delta \epsilon
}{\epsilon} \approx C_{\epsilon} \frac{\Delta V_{0\bot }}{V_{0\bot}}. 
\end{equation}
with the susceptibilities $C_J$, $C_U$, and $C_\epsilon$ of the respective Bose-Hubbard parameter. They depend on the working point $V_{0\bot}$ and are given in \tab{tab:chi} for $^{87}$Rb.

\begin{table}
\begin{ruledtabular}
\begin{tabular}{ll|r|r|r|r}
Parameter &	& $^7$Li	 & $^{23}$Na & $^{41}$K & $^{87}$Rb \\
\hline
$J/h$ & [Hz] & $18.6$ & $3.5$ & $2.3$ & $1.5$\\
$U/h$ & [Hz] & 
$186$ & $35$ & $23$ & $15$\\
$V_{0\bot}/k_B$ & [nK] & $2253$ & $786$ & $425$ & $181$\\
$V_{0\parallel}/k_B$ & [nK] & $4971$ & $1736$ & $939$ & $400$\\
$\tau_\text{ramp}$ & [ms] & $62$ & $330$ & $502$ & $770$ \\
$\Gamma_\text{sc}$ & [s$^{-1}$] & $0.01$ & $0.003$ & $0.004$ & $0.002$\\
$N_\text{sc}$ &  & $0.05$ & $0.09$ & $0.18$ & $0.14$\\
\end{tabular}
\end{ruledtabular}
\caption{Parameters for different atomic species. Experimental parameters of a square lattice for various bosonic alkalies at $U/J=10$ (typical for the superfluid/Mott-insulator transition): tunneling energy $J$, on-site interaction energy $U$, in-plane and out-of-plane potential depths $V_{0\bot}$ and $V_{0\parallel}$, adiabatic ramp time $\tau_\text{ramp}$, scattering rate per atom $\Gamma_\text{sc}$, and total number of scattering events $N_\text{sc}$ for a lattice of $N=100$ atoms  during the loading process. For the computation of $U$ it is assumed that the atoms are prepared in the state $|F=1,m_F=-1\rangle$. For the case of $^{7}$Li that the scattering length is tuned to \unit{5.3}{nm} via a Feshbach resonance.}
\label{table}
\end{table}

We estimate the experimental fluctuations of the trap depths by assuming that the total power of the laser illuminating the SLM can be stabilized to a relative uncertainty of $0.1 \%$. Further, we presume that the illumination of the MLA is controlled by a liquid crystal based SLM with $768$ pixel per dimension combined with a polarizing beam splitter. The transmission of each pixel can be controlled in $256$ steps \cite{Kruse2010}. For a square MLA consisting of $25 \times 25$ lenses with circular aperture, $741$ pixels per lens are available. This can be used to compensate static imperfections arising from production tolerances, aberrations, or an inhomogeneous beam profile on the $\Delta V_{0\bot}/V_{0\bot} = (741 \cdot 256)^{-1}=5.3 \cdot 10^{-6}$ level.

In order to allow for tunneling between adjacent traps, the difference between the single particle energies $\Delta \epsilon$ must be smaller than the tunneling parameter $J$. Due to the small ratio of $\epsilon /J$ this was found to be challenging in experiments with optical tweezers \cite{Kaufman2014, Murmann2015}. Our estimate for the spatial fluctuations of the trap depth results in $\Delta\epsilon/J = 1 \%$ at $U/J=10$. Comparing this value to the parameters of experiments with double-well configurations \cite{Kaufman2014, Murmann2015} shows that unobstructed tunneling is feasible in our setup. Temporal fluctuations in the laser power result in global changes in the potential depth and are therefore not relevant for this aspect. 

Measurements rely on averages over repeated experimental runs. Therefore, in order to resolve structures like the many-body resonances shown in \fig{fig:CoupledRingsResonances} (a) spatial and temporal fluctuations need to be small compared to the respective structure's width. For the sharp resonance at $U/J=5.07$, we find a theoretical full width at half maximum of $0.06 \; U/J$. Using $\Delta V_{0\bot}/V_{0\bot} = 0.1 \%$ for the expected experimental fluctuations and the linear susceptibilities from table \ref{tab:chi}, we obtain an expected uncertainty of $\Delta (U/J)= 0.015$. Therefore, even the sharpest resonance in \fig{fig:CoupledRingsResonances} (a) can be resolved.

\subsection{Light scattering}

The heating induced by the loading procedure \cite{McKay2011} is a serious concern. Typically, low-entropy states are produced by preparing a BEC in a large scale harmonic trap and subsequently ramping up the lattice potential. Sufficiently long ramp times $\tau_\text{ramp}$ will lead to an adiabatic loading process conserving the entropy of the initial state. To estimate the required $\tau_\text{ramp}$, we scale the experimental result of reference \cite{Sherson2010} according to the lattice spacing $d=\unit{1.7}{\micro m}$. The results are given in table \ref{table} and range between  $\tau_\text{ramp}=\unit{62}{ms}$ for $^7$Li and $\tau_\text{ramp}=\unit{770}{ms}$ for $^{87}$Rb. However, considerably faster ramps might be possible in our platform since it facilitates homogeneous lattice potentials omitting transport through Mott phases which limits ramp speeds in conventional experiments \cite{Dolfi2015}.

Finally, the scattering rate of photons $\Gamma_\text{sc}$ from the light field generating the optical potential has to be considered. This is the major source of heating in optical lattice experiments \cite{McKay2011}. To estimate this effect, we consider the impact of a single scattering event and their total number during the adiabatic loading process lasting $\tau_{\text{ramp}}$. The total number of scattering events $N_\text{sc}=N\int_0^{\tau_\text{ramp}} \Gamma_\text{sc} (t) dt$, scales with the number of atoms $N$. We consider a $10\times10$ lattice filled with one atom per site, giving $N=100$. Table~\ref{table} confirms that for a linear ramp with length $\tau_\text{ramp}$ and final lattice parameters $V_{0\perp}$ and $V_{0\parallel}$ hardly any scattering event occurs. The impact of a single scattering event can be estimated to be ten times more severe in comparison to traditional optical lattices, however the low overall number of events renders the adverse effect negligible, nevertheless.

\begin{table}
\centering
\begin{ruledtabular}
\begin{tabular}{c|c|c|c|c}
$V_{0\bot}/k_B$ (nK) &$U/J$ & $C_J$ & $C_U$ & $C_\epsilon$ \\
\hline
$89.5$ &1 & $1.32\pm 0.07$ & $0.87 \pm 0.04$ & $1.49\pm 0.06$ \\
$150.5$ &5 & $2.09\pm 0.04$ & $0.72\pm 0.03$ & $1.36\pm 0.05$ \\
$181.1$ &10 & $2.43\pm 0.10$ & $0.67\pm 0.02$ & $1.29\pm 0.04$ \\
$200.3$ &15 & $2.58\pm 0.09$ & $0.68\pm 0.03$ & $1.32 \pm 0.05$ \\
$225.7$ &25 & $2.78\pm 0.11$ & $0.66\pm 0.02$ & $1.29\pm 0.04$ \\
\end{tabular}
\end{ruledtabular}
\caption{Linear susceptibilities of Bose-Hubbard parameters versus mean potential depth $V_{0\bot}$ for $^{87}$Rb. The uncertainties are $95 \%$ confidence intervals of the linear fit.}
\label{tab:chi}
\end{table}

\section{Conclusions and Outlook}
\label{sec:outlook}
In conclusion, we have introduced and analyzed a novel experimental platform for a freely configurable quantum simulator for many-body physics using ultra-cold atoms in lattice-type geometries of focused beam dipole traps. With a large lattice spacing ($d=\unit{1.7}{\micro m}$) this setup enables dynamic individual-site control of each potential well, every trapped atom and individual interaction strength without impact on neighboring sites. Simulations based on measured optical intensities prove the feasibility of the cross-over to the strongly interacting many-body regime. As a first application, we have designed a weak Josephson link of two ring lattices. The analysis of a Bose-Hubbard model exhibits interesting many-body resonances for the enhancement of population transfer in the strongly interacting regime.

In the future, even more advanced options of our scheme can be foreseen: 1) The setup's dynamic single-site control can be exploited to lower the entropy of the final many-body state by generating an inhomogeneous envelope for the lattice potential resulting in the coexistence of spatially separated Mott-insulator and superfluid phases. The superfluid part carries most of the systems entropy and can be removed by emptying the respective traps \cite{Bernier2009,McKay2011}. The resulting low entropy state opens a route to cold-atom analogs of high-T$_c$ superconductors \cite{McKay2011}. 2) The dynamic control over local potential depths can be utilized for Floquet engineering \cite{Meinert2016, Goldman2014} with modulation frequencies of up to \unit{10}{kHz}, fully adjustable modulation amplitudes, and single-site addressability. 3) A promising alternative to the loading schemes starting with BECs arises from the implementation of Raman side-band cooling in individual traps \cite{Kaufman2012, Thompson2013} with the targeted many-body state assembled atom by atom \cite{Barredo2016, Endres2016, Kim2016} out of the low entropy Mott-insulator phase. This facilitates studies of the many-body physics of atomic species, which are  not accessible to BEC, or arbitrary mixtures of species.

\section*{Acknowledgments}

We thank J.-N. Schmidt for measuring the experimental intensity distributions. M.R.S. and R.W. acknowledge support from the German Aeronautics and Space Administration (DLR) through grant 50 WM 1557. M.S. and G.B. acknowledge support by the Deutsche Forschungsgemeinschaft (DFG) through grant BI647/6-1 within the Priority Program SPP 1929 (GiRyd).

\appendix

\section{Weak interaction limit}
\label{sec:appendix}
Here, we discuss the regime of weak interactions for the system discussed in section \ref{sec:rings}. For the non-interacting system ($U=K=0$), the ground state manifold $\mathcal{G}_0$ is $N+1$-fold degenerate since the distribution of the particles between the rings has no effect on the energy. The eigenstates 
\begin{equation}
| n \rangle=
\frac{
%\hat{\underline{a}}^{\dagger n}_0 
%\hat{\underline{b}}^{\dagger N-n}_0 
\ald{n}{0}
\betd{N-n}{0}
}{\sqrt{n!(N-n)!}} 
|0\rangle,\quad
\al{}{l} = \sum\limits_{m=0}^{M-1} \frac{e^{-2 \pi \jmath \frac{lm}{M}}}{\sqrt{M}} 
\aop{m},
\end{equation}
are labeled by $0\le n \le N$ and are defined by the Fourier operators $\al{}{l}$ for ring A and $\bet{}{l}$ for ring B with $0\leq l<M$. For $NU,K\ll J$, only states within $\mathcal{G}_0$ contribute to the system's dynamics. By restricting the dynamics to the two lowest Fourier amplitudes, the Hamilton operator reads
\begin{equation}
\label{h00}
\hat{H}_0= 
-k (
\ald{}{0} \bet{}{0} 
+ 
\betd{}{0}
\al{}{0} )
+ 
\frac{u}{2} (
\ald{}{0}
\ald{}{0}
\al{}{0}
\al{}{0}
+
\betd{}{0}
\betd{}{0}
\bet{}{0}
\bet{}{0}
)
% -2JN.
\end{equation}
up to an additive constant. This is an effective two-site Bose-Hubbard model with tunneling strength $k=K/M$ and interaction strength $u=U/M$. For $u=0$, the canonical transformation $\hat{c}_\pm = (\al{}{0} \pm \bet{}{0})/\sqrt{2}$ diagonalizes (\ref{h00}) and the eigenstates read 
\begin{equation}
\label{basis}
|n) = \frac{\hat{c}_-^{\dagger n} \hat{c}_+^{\dagger N-n}}{\sqrt{n!(N-n)!}}  
|0\rangle.
\end{equation}
To first order in $u$, the energy $E_n=(n|\hat{H}_0|n)$ reads
\begin{equation}
E_n = k (2n-N)+
\frac{u}{4}[N(N-1) + 
2n(n - N)].
\end{equation}
If all $N$ atoms are initially in ring $A$, i.\thinspace{}e., $|\psi(0)\rangle=|N\rangle$, it evolves approximately as
\begin{equation}
|\psi (t)\rangle = 
\frac{1}{2^{N/2}}
\sum\limits_{n=0}^N 
\begin{pmatrix}N\\n\end{pmatrix}^{1/2}
%\sqrt{\frac{N!}{2^N n! (N-n)!}} 
e^{-\jmath E_n t/\hbar}|n)
\end{equation}
introducing  the binomial coefficient. Hence, the population inversion
\begin{align}
\zeta(t)&=
\frac{1}{2^{N-1}}\sum\limits_{n=0}^N  
\begin{pmatrix}N-1\\n\end{pmatrix}
\cos{( \frac{E_{n+1}-E_n}{\hbar}t)}\nonumber\\
&=\cos(\omega_j t) \cos(\pi t/\tau_r)^{N-1}.
\end{align}
follows (cf. Eq. (3) in the main paper). A similar analysis has been performed for the two-site Bose-Hubbard model in the limit of large $N$ where the collapse is predicted to have a Gaussian shape \cite{Veksler2015}.

%\bibliography{references}

%

\end{document}